\documentclass[10pt]{article}
\usepackage{geometry}
\usepackage[utf8x]{inputenc}
\usepackage{graphicx}
\usepackage{amsmath}
\usepackage{ulem}

\def\bd{\begin{displaymath}}\def\ed{\end{displaymath}}
\def\be{\begin{equation}}\def\ee{\end{equation}}
\def\bea{\begin{eqnarray}}\def\eea{\end{eqnarray}}
\def\ba{\begin{array}}\def\ea{\end{array}}
\def\lb{\label}

\def\b{\beta}\def\d{\delta}

\def\k{\kappa}\def\l{\lambda}\def\m{\mu}\def\n{\nu}
\def\y{\eta}\def\x{\xi}

\def\D{\Delta}

\def\de{\partial}
\def\inf{\infty}\def\ha{{1\over 2}}

\def\bdot{\!\cdot\!}\def\mn{{\mu\nu}}

\def\coo{coordinates }\def\eom{equations of motion }\def\ie{i.e.\ }
\def\bc{boundary conditions }\def\trans{transformations }
\def\kp{$\k$-Poincar\'e }

\def\PL#1{Phys.\ Lett.\ {\bf#1}}

\def\PR#1{Phys.\ Rev.\ {\bf#1}}

\def\JoP#1{J.\ Phys.\ {\bf#1}} \def\IJMP#1{Int.\ J. Mod.\ Phys.\ {\bf #1}}

\def\JHEP#1{JHEP\ {\bf#1}}

\def\cK{{\cal K}}\def\zero{{[0]}}\def\uno{{[1]}}

\begin{document}

\title{Relative-locality phenomenology on Snyder spacetime}

\author{$~$\\
{\bf Salvatore Mignemi}$^{1,2}$, {\bf Giacomo Rosati}$^{2}$
\\
$^1${\footnotesize Dipartimento di Matematica e Informatica, Universit\`a di Cagliari, viale Merello 92, 09123 Cagliari, Italy }\\
$^2${\footnotesize  INFN, Sezione di Cagliari, Cittadella Universitaria, 09042 Monserrato, Italy }}
\maketitle

\begin{abstract}
We study the effects of relative locality dynamics in the case of the Snyder model.
Several properties of this model differ from those of the widely studied \kp models:
for example, in the Snyder case the action of the Lorentz group is preserved, and the 
composition law of momenta is deformed by terms quadratic in the inverse Planck energy.
From the investigation of time delay and dual curvature lensing we deduce that, because
of these differences, in the Snyder case the properties of the detector are essential
for the observation of relative locality effects. The deviations from special relativity
do not depend on the energy of the particles and are much smaller
than in the \kp case, so that are beyond the reach of present astrophysical experiments.
However, these results have a conceptual interest, because they show that relative-locality
effects can occur even if the action of the Lorentz group on phase space is not deformed.
\end{abstract}

\newpage
\section{Introduction}

The possibility of testing quantum spacetime scenarios via Planck scale effects in the kinematics of point-like particles became of primary relevance for quantum gravity phenomenology over the last 15 years~\cite{GACphen,MattinglyRev,GACGRB,AlfaroNeutrinos,JacobPiran,IceCubeNat,GACJrFioreGRB}.
In this perspective the ``doubly special relativity'' (or, as some authors prefer to call it, the ``deformed relativistic symmetries) (DSR) scenario~\cite{GACDSR,JurekDSR,SmolMagDSR}, in which the Planck scale enters in the laws of motion as an observer-independent scale deforming the algebra of relativistic symmetries, has played a prominent role (see for instance~\cite{DSRFRW,DualLensing}).

As the understanding of DSR advanced it has been realized that the introduction in a relativistic theory of a second invariant scale (in addition to the speed of light)  with dimensions of an inverse momentum, i.e. proportional to the inverse of the Planck energy $E_p\sim 10^{19} \text{GeV}$,  enforces to abandon the concept of absoluteness of locality: the locality of a process is relative to the distance from the observer, in such a way that a process remains local only for an observer local to the process itself~\cite{RelLocPRL,kBob,DSRFRW}.
The relative locality framework~\cite{RelLocPrinciple} (see also~\cite{CarmonaRelLoc,JurekReview,Palmisano}) has been proposed as a formulation of DSR capable of taking into account relative locality effects, focusing on the (curved) momentum space associated to the quantum spacetime deformation, and taking into account the kinematical properties of particle processes by means of a Lagrangian formulation with suitable boundary terms.

So far the relative locality framework has been investigated extensively only for the well-known case of $\kappa$-Poincaré symmetries~\cite{anatomy,FlaGiukRelLoc,causality,multipart}, while some work has been done for a model of momentum space based on a quantum spacetime with non-commutativity of SL(2,R) type, called ``spinning spacetime'' by the authors in~\cite{spinning}.
The purpose of the present paper is to study the implications of the relative locality framework for another well-known model of deformed momentum space, based on a quantum spacetime with Snyder type non-commutativity~\cite{Snyder,Sn,BM}.
The peculiarity of this model is the preservation of the linear action of the Lorentz group on phase space.
This can help to single out the role of the deformation of Lorentz invariance on the effects connected to relative locality.
In fact, it has been shown that no such effects are present in the case of the propagation of free particles \cite{MigSamSnyderRL} (see also~\cite{AstutiSnyder}).
However, when considering interacting particles, one must take into account that the composition law of momenta is deformed \cite{BM,IMS},
and hence one can expect relative-locality effects in more complicated settings involving interactions.
However, these are presumably greatly suppressed and far from the reach of present experimental sensibility.
It is important to notice indeed that while the $\kappa$-deformation implies modifications to kinematical laws of linear order in $1/E_p$, in Snyder momentum space the deformation scale is proportional to the square of the inverse Planck energy, $\beta\propto 1/E_p^2$. We thus expect that the possible phenomenological outcomes, if any, should be suppressed by a further Planck-scale factor with respect to the the already tiny ones arising from
$\kappa$-deformation.
However, it is interesting at least from a theoretical point of view to show the existence of some effects associated with Snyder spacetime arising from the relative locality formalism.
We will devote the last part of this manuscript to the study of some relevant processes that could provide this kind of effects.

Another peculiarity of Snyder spaces is that the composition law of momenta is not only noncommutative, as in most models of DSR,
but also nonassociative \cite{BM}, determining further complications in the investigation of interactions.
To some extent, at the classical level, the nonassociativity is less disturbing than at the quantum level,
since a natural ordering can be assumed, at least for three-particle interactions.
Still, when one considers more than one process, as for causally connected interactions, a further ambiguity arises in the choice of how to group summations of more than two momenta,
in addition to the one due to noncommutativity.
We will discuss this issue in the course of our analysis.

Finally, we recall that the geometrical properties of relative locality momentum space introduced in ref.~\cite{RelLocPrinciple} have been investigated in \cite{IM} in the case of Snyder space and its generalizations.
It turns out that the momentum space associated to these models is a maximally symmetric one, with constant curvature and vanishing torsion and nonmetricity. However, for our investigations these results are irrelevant, so we refer to \cite{IM}  the interested reader.

Throughout the manuscript we use units such that the speed of light $c$ is set to 1.

\section{The Snyder model}
The Snyder model \cite{Snyder,Sn,BM} is defined by the deformed Heisenberg algebra
\be\lb{sny}
\{x^\m,p_\n\}=\d^\m_\n-\b p^\m p_\n,\qquad\{x^\m,x^\n\}=-\b J^\mn,\qquad\{p_\m,p_\n\}=0,
\ee
where $J^\mn=x^\m p^\n-p^\m x^\n$ are the generators of the Lorentz transformations.
The parameter $\b$ has dimension of inverse mass square and is usually assumed to be of order $1/E_p^2$.
The Lorentz algebra and its action on phase space are undeformed.
Snyder space can be also described in terms of a curved momentum space given by a hyperboloid of equation
$\y_A^2=-1/\b$ embedded in flat five-dimensional space of \coo $\y_A$, with parametrization $p_\m=\y_\m/\sqrt\b\y_4$.

Since the Lorentz transformations are undeformed, the dispersion relations maintain the same form as in special relativity, and the
Hamiltonian of a free particle can be chosen as\footnote{We denote $a^\m b_\m\equiv a\bdot b$.}
\be\lb{ham}
H={p\bdot p\over2}.
\ee
The \eom following from (\ref{sny}) and (\ref{ham}) are
\be\lb{eomp}
\dot x^\m=\{x^\m,H\}=(1-\b p\bdot p)p^\m,\qquad\dot p_\m=\{p_\m,H\}=0.
\ee
They can be obtained varying the action \cite{MS}
\be\lb{act}
S=-\int_{-\inf}^{\inf}  ds\left[x^\m\left(\y_\mn+\b{p_\m p_\n\over1-\b p\bdot p}\right)\dot p^\n+{N\over2}\,(p\bdot p-m^2)\right].
\ee
where $N$ is a Lagrange multiplier enforcing the Hamiltonian constraint $p\cdot p=m^2$.

Starting from (\ref{sny}), one can define a Hopf algebra \cite{BM}. The coproduct of this algebra entails the
deformed addition law for momenta,
\be
(p\oplus q)_\mu={1\over 1+\b p\bdot q}\left[\left(1+{\b p\bdot q\over1+\sqrt{1-\b p\bdot p}}\right)p_\m+\sqrt{1-\b p\bdot p}\ q_\m\right].
\label{Oplus}
\ee
Notice that this law is noncommutative, $p\oplus q\ne q\oplus p$, as in other well-known models, but also nonassociative,
$k\oplus(p\oplus q)\ne(k\oplus p)\oplus q$.

The antipode of the element $p$ of the Hopf algebra, \ie the element $\ominus p$ such that $p\oplus(\ominus p)=0$,
is given by
\be\lb{ant}
\ominus p=-p.
\ee
\begin{figure}[h!]
\centering
\includegraphics[scale=0.7]{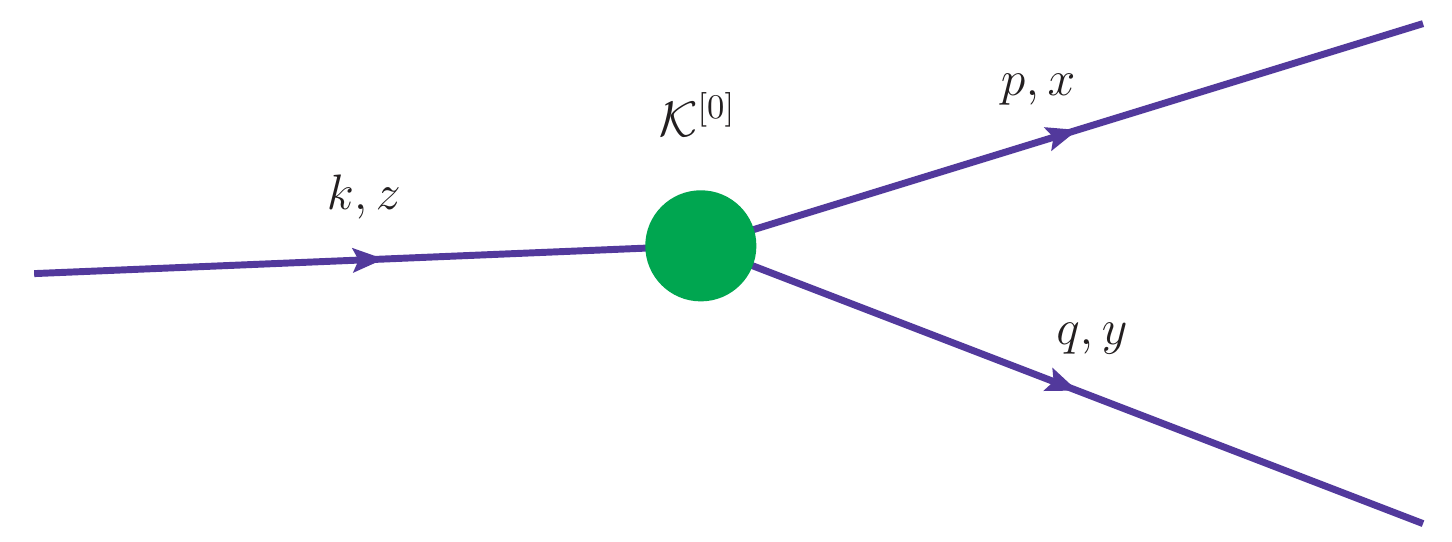}\caption{\small Illustrative representation of a trivalent vertex at $s_0$ with an incoming particle $(k,z)$ and two outgoing particles $(p,x)$ and $(q,y)$.\label{fig:Snyder1vertex}}
\end{figure}
\section{Interactions in relative locality}
We start now the discussion of the dynamics on Snyder space in accordance with the relativistic description of distant
observers given by the framework of relative locality \cite{RelLocPrinciple}. The action for a noninteracting particle is of course given
by (\ref{act}). The definition of an action for interacting particles requires instead some discussion.
In this section, we consider a simple interaction with one incoming and two outgoing particles,
and use a Lagrangian formalism, following the treatment given in \cite{anatomy} for the case of the \kp model.

In this formalism, the action for three free particles of momenta $k_\m$, $p_\m$, $q_\m$
and masses $m_k$, $m_p$, $m_q$, with positions $z^\m$, $x^\m$, $y^\m$, respectively, interacting at parameter time
$s_0$  (see Fig.~\ref{fig:Snyder1vertex}), can be written adding to the terms describing the propagation of the noninteracting particles an
interaction term $-\x_\zero^\m\cK^\zero_\m$ \cite{RelLocPrinciple}, as
\bea\lb{intact}
S&=&-\int_{-\inf}^{s_0}ds\left[\left(z^\m+\b{z\bdot k\,k^\m\over1-\b k\bdot k}\right)\dot k_\m+{N_k\over2}(k\bdot k-m_k^2)\right]\cr
&&-\int_{s_0}^\inf ds\left[\left(x^\m+\b{x\bdot p\,p^\m\over1-\b p\bdot p}\right)\dot p_\m+{N_p\over2}(p\bdot p-m_p^2)\right]\cr
&&-\int_{s_0}^\inf ds\left[\left(y^\m+\b{y\bdot q\,q^\m\over1-\b q\bdot q}\right)\dot q_\m+{N_q\over2}(q\bdot q-m_q^2)\right]
+\x_\zero^\m\cK^\zero_\m,
\eea
where $\cK^\zero_\m=0$ is the conservation law at the interaction, while $N_k$, $N_p$, $N_q$ and
$\x^\m_\zero$ are Lagrange multipliers. However, the $\x^\m_\zero$  can be interpreted as interaction coordinates,
which vanish for an observer local to the interaction, but not for distant observers \cite{RelLocPrinciple}.
The \coo $z^\m$, $x^\m$, $y^\m$ are instead the position \coo measured by generic observers. This interpretation
is enforced by the crucial requirement that translations of parameter $b^\m$ are generated by the momentum conservation
law $\cK^\zero_\m$,\footnote{This prescription is not standard, however for undeformed translations, it is equivalent
to the usual one that the translations are generated by the momentum of the particle.}
and then identifying $b^\m$ with $\x_\zero^\m$ \cite{RelLocPrinciple,anatomy}. From the boundary
equations (\ref{bc}) below, we can in fact see that $z^\m$, $x^\m$ and $y^\m$ all vanish when $\x^\m_\zero=0$,
but in general they are different when $\x^\m_\zero\ne0$.
This is the principle of relative locality: interactions that are local for an observer at the interaction point
are seen by distant observers as nonlocal.

Several possibilities exist for the definition of $\cK^\zero_\m$. The most natural one,
$\cK^\zero_\m=k_\m\ominus(p\oplus q)_\m$, does not satisfy the relativity requirement for the \kp model \cite{anatomy}. The same
problem arises for the Snyder model. The problem emerges when one considers finite worldlines with two endpoints.
In this case, requiring that the \eom for two observers at rest $A$ and $B$ be the same, as required by the relativity
principle \cite{RelLocPrinciple}, implies that the translations generated by the 'total momentum' (conservation laws) at the two ends of
the worldlines give
rise to the same relation between the position measured by the two observer, and in particular that
\be\lb{rel}
{\de\cK^\zero_\m\over\de p_\m}=-{\de\cK^\uno_\m\over\de p_\m},
\ee
where $\cK^\uno_\m$ is the conservation law at the second end of the worldline (see sect.~4 for more details).
With the previous definition of $\cK^\zero_\m$ it is not possible to satisfy this request. However, because of (\ref{ant}),
the conservation law can be equivalently written as
\be\lb{cons}
\cK^\zero_\m=k_\m-(p\oplus q)_\m=0,
\ee
and this expression permits to overcome the problem, as it was observed in \cite{anatomy} for the case of the \kp model.
This form of the conservation law expresses the requirement that the total momentum $k_\m$ before the interaction  be equal
to the one after the interaction, $(p\oplus q)_\m$.
In particular, in our case we define
\be
\cK^\zero_\m=k_\m-{1\over 1+\b p\bdot q}\left[\left(1+{\b p\bdot q\over1+\sqrt{1-\b p\bdot p}}\right)p_\m+\sqrt{1-\b p\bdot p}\ q_\m\right]
\ee

Note that this expression, due to the noncommutativity of the addition law of momenta, is not symmetric under the
exchange of the outgoing particles. This feature has been already discussed in~\cite{anatomy}, where it has been shown that different
orderings for the momenta appearing in the boundary terms produce in general different predictions for the physical observables.
One can interpret the different ordering choices as different channels for the process.
Moreover,  due to the nonassociativity of the addition law, (\ref{cons})  is also not invariant if one changes the grouping of the
momenta in the sum. However, at the classical level it seems natural to choose an expression like (\ref{cons})
that distinguishes the incoming particles from the outgoing ones. At the quantum level, or in more complicated processes, one may
however be forced to consider also different orderings of the sums (see sect.~4).

For simplicity, from now on we consider the linearized theory, although it is not difficult to reproduce the calculation
in the nonlinear case. The linearization of the action (\ref{intact}) yields
\bea\lb{lin}
S&=&-\int_{-\inf}^{s_0}ds\left[(z^\m+\b z\bdot k\,k^\m)\dot k_\m+{N_k\over2}(k\bdot k-m_k^2)\right]\cr
&&-\int_{s_0}^\inf ds\left[(x^\m+\b x\bdot p\,p^\m)\dot p_\m+{N_p\over2}(p\bdot p-m_p^2)\right]\cr
&&-\int_{s_0}^\inf ds\left[(y^\m+\b y\bdot q\,q^\m)\dot q_\m+{N_q\over2}(q\bdot q-m_q^2)\right]+\x_\zero^\m\cK^\zero_\m+O(\b^2),
\eea
with
\be
\cK^\zero_\m=k_\m-p_\m-q_\m+{\b\over2}\big(p\bdot q\,p_\m+2p\bdot q\,q_\m+p\bdot p\,q_\m\big)+O(\b^2)
\ee
The \eom derived from (\ref{lin}) are
\bea\lb{eom}
&&\dot k_\m=\dot p_\m=\dot q_\m=0,\qquad\qquad k\bdot k-m_k^2=p\bdot p-m_p^2=q\bdot q-m_q^2=0,\cr
&&\dot z_\m=N_k(1-\b k\bdot k)k_\m,\quad\dot x_\m=N_p(1-\b p\bdot p)p_\m,\quad\dot y_\m=N_q(1-\b q\bdot q)q_\m,
\eea
together with $\cK^\zero_\m(s_0)=0$.

The boundary terms at $s=s_0$ yield
\bea\lb{bond}
&&(\y_\mn+\b k_\m k_\n)\,z^\n(s_0)={\de\cK^\zero_\n\over\de k_\m}\,\x_\zero^\n=\x_\zero^\m,\cr
&&(\y_\mn+\b p_\m p_\n)\,x^\n(s_0)={\de\cK^\zero_\n\over\de p_\m}\,\x_\zero^\n,\cr
&&(\y_\mn+\b q_\m q_\n)\,y^\n(s_0)={\de\cK^\zero_\n\over\de q_\m}\,\x_\zero^\n,
\eea
where
\bea
&&{\de\cK^\zero_\n\over\de p_\m}\x_\zero^\n=-\x_\zero^\m+\b\left(q^\m q_\n+p^\m q_\n+\ha p_\n q^\m+\ha p\bdot q\,\d^\m_\n\right)\,\x_\zero^\n,\cr
&&{\de\cK^\zero_\n\over\de q_\m}\x_\zero^\n=-\x_\zero^\m+\b\left(\ha p^\m p_\n+p^\m q_\n+\ha p\bdot p\,\d^\m_\n+p\bdot q\,\d^\m_\n\right)\,\x_\zero^\n.
\eea
Inverting the matrices at the left hand side, (\ref{bond}) can be put in the form
\bea\lb{bc}
&&z^\m(s_0)=(\d^\m_\n-\b k^\m k_\n)\,\x_\zero^\n,\cr
&&x^\m(s_0)=(\d^\m_\n-\b p^\m p_\n){\de\cK^\zero_\l\over\de p_\n}\,\x_\zero^\l,\cr
&&y^\m(s_0)=(\d^\m_\n-\b q^\m q_\n){\de\cK^\zero_\l\over\de q_\n}\,\x_\zero^\l.
\eea
As discussed before, the \bc establish that if the observer is local to the interaction, \ie $\x_\zero^\m=0$, the endpoints
of the worldlines of all the particles are at the origin of the observer, while if $\x_\zero^\m\ne0$, the endpoints of the
worldlines of different particles do not coincide.
Under infinitesimal deformed translations of parameter $b^\n$ between two observers $A$ and $B$, generated by the total
momentum $\cK^\zero_\n$, the spacetime \coo transform as
\bea\lb{trans}
&&z_B^\m(s)=z_A^\m+b^\n\{\cK^\zero_\n,z^\m\}=z_A^\m(s)-(\d^\m_\n-\b k^\m k_\n)b^\n,\cr
&&x_B^\m(s)=x_A^\m+b^\n\{\cK^\zero_\n,x^\m\}=x_A^\m(s)-(\d^\m_\n-\b p^\m p_\n){\de\cK^\zero_\l\over\de p_\n}\,b^\l,\cr
&&y_B^\m(s)=y_A^\m+b^\n\{\cK^\zero_\n,y^\m\}=y_A^\m(s)-(\d^\m_\n-\b q^\m q_\n){\de\cK^\zero_\l\over\de q_\n}\,b^\l.
\eea
The \eom and the \bc are invariant under these \trans if
\be\lb{cond}
\x_{\zero B}^\m-\x_{\zero A}^\m=b^\m.
\ee
In fact, the \eom (\ref{eom}) only contain the momenta and the time derivatives of the coordinates. The momenta are obviously
invariant under translations, while differentiating (\ref{trans}) one sees that also $\dot z$, $\dot x$ and $\dot y$ are invariant,
because the momenta are conserved. Moreover, using (\ref{trans})  it is easy to check that the \bc(\ref{bc}) are invariant if (\ref{cond})
holds.


Of course one could choose a different ordering for the outgoing momenta, $\cK^\zero_\m=k_\m-(q\oplus p)_\m$. Although the
momenta $p$ and $q$ are interchanged everywhere, the main conclusions are unaltered. As mentioned above, one may interpret this
fact as the existence of two different channels for the interaction.

\section{Causally connected interactions}
\begin{figure}[h!]
\centering
\includegraphics[scale=0.7]{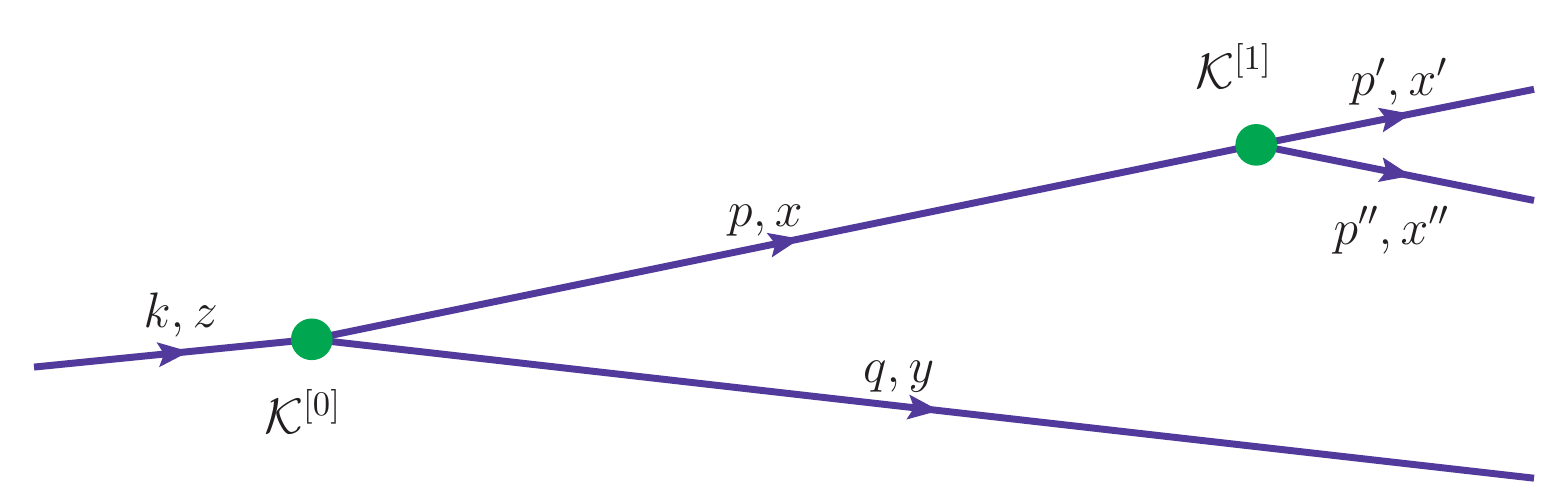}\caption{\small In this schematic representation the two vertices at $s_0$ and $s_1$ are causally connected by a particle produced at $s_0$ in a
trivalent interaction and having a second trivalent interaction at $s_1$.\label{fig:Snyder2vertex}}
\end{figure}

We can now consider two causally connected interactions, occurring at $s_0$ and $s_1$, as depicted in Fig.~\ref{fig:Snyder2vertex}.

The linearized action is the obvious generalization of (\ref{lin}):
\bea
&&S=-\int_{-\inf}^{s_0}ds\left[(z^\m+\b z\bdot k\,k^\m)\dot k_\m+{N_k\over2}(k\bdot k-m_k^2)\right]
-\int_{s_0}^{s_1} ds\left[(x^\m+\b x\bdot p\,p^\m)\dot p_\m+{N_p\over2}(p\bdot p-m_p^2)\right]\cr
&&-\int_{s_0}^\inf ds\left[(y^\m+\b y\bdot q\,q^\m)\dot q_\m+{N_q\over2}(q\bdot q-m_q^2)\right]
-\int_{s_1}^\inf ds\left[(x'^\m+\b x'\bdot p'\,p'^\m)\dot p'_\m+{N_{p'}\over2}(p'\bdot p'-m_{p'}^2)\right]&\cr
&&-\int_{s_1}^\inf ds\left[(x''^\m+\b x''\bdot p''\,p''^\m)\dot p''_\m+{N_{p''}\over2}(p''\bdot p''-m_{p''}^2)\right]
+\x^\m_\zero\cK^\zero_\m+\x^\m_\uno\cK^\uno_\m.
\label{actionCausal}
\eea
The \eom read
\bea
&&\dot k_\m=\dot p_\m=\dot q_\m=\dot p'_\m=\dot p''_\m=0,\cr
&&k\bdot k-m_k^2=p\bdot p-m_p^2=q\bdot q-m_q^2=p'\bdot p'-m_{p'}^2=p''\bdot p''-m_{p''}^2=0,\cr
&&\dot z_\m=N_k(1-\b k\bdot k)k_\m,\quad\dot x_\m=N_p(1-\b p\bdot p)p_\m,\quad\dot y_\m=N_q(1-\b q\bdot q)q_\m,\cr
&&\dot x'_\m=N_{p'}(1-\b p'\bdot p')p'_\m,\quad\dot x''_\m=N_{p''}(1-\b p''\bdot p'')p''_\m,
\eea
together with $\cK^\zero_\m(s_0)=\cK^\uno_\m(s_1)=0$.

The \bc at $s_0$ still yield (\ref{bc}), while those at $s_1$ give
\bea
&&x^\m(s_1)=(\d^\m_\n-\b p^\m p_\n){\de\cK^\uno_\l\over\de p_\n}\,\x_\uno^\l\cr
&&x'^\n(s_1)=(\d^\m_\n-\b p'^\m p'_\n){\de\cK^\uno_\l\over\de p'_\n}\,\x_\uno^\l\cr
&&x''^\n(s_1)=(\d^\m_\n-\b p''^\m p''_\n){\de\cK^\uno_\l\over\de p''_\n}\,\x_\uno^\l.
\eea

Under infinitesimal translation generated by $\cK^\uno_\n$, one has
\bea\lb{trans2}
&&x_B^\m=x_A^\m+b^\n\{\cK^\uno_\n,x^\m\}=x_A^\m-(\d^\m_\n-\b p^\m p_\n){\de\cK^\uno_\l\over\de p_\n}\,b^\l,\cr
&&x'^\m_B=x'^\m_A+b^\n\{\cK^\uno_\n,x'^\m\}=x'^\m_A-(\d^\m_\n-\b p'^\m p'_\n){\de\cK^\uno_\l\over\de p'_\n}\,b^\l\cr
&&x''^\m_B=x''^\m_A+b^\n\{\cK^\uno_\n,x''^\m\}=x''^\m_A-(\d^\m_\n-\b p''^\m p''_\n){\de\cK^\uno_\l\over\de p''_\n}\,b^\l.
\eea

As mentioned before, a crucial requirement for the choice of the conservation laws at the interactions
is that in the case of multiple interactions the relation (\ref{rel}) holds.
This is necessary in order to get the same \eom for both observers $A$ and $B$: in fact,
\be\lb{bc2}
x_B^\m(s_0)=x_A^\m(s_0)+b^\l(\d^\m_\n-\b p^\m p_\n){\de\cK^\zero_\l\over\de p_\n},
\qquad x_B^\m(s_1)=x_A^\m(s_1)-b^\l(\d^\m_\n-\b p^\m p_\n){\de\cK^\uno_\l\over\de p_\n},
\ee
and $\dot x_\m=N_p(1-\b p\bdot p)p_\m$ can hold for both observers only if (\ref{rel}) is verified.
This implies that the translation parameters do not depend on $s$.
In order to satisfy (\ref{rel}), the authors of \cite{anatomy} propose for $\cK^\uno$, in the case of \kp
\be
\cK^\uno=(p\oplus q)-(p'\oplus p''\oplus q).
\ee

In our case, the second expression is not well defined, because of nonassociativity,
but one may set
\be\lb{cl}
\cK^\uno=(p\oplus q)-((p'\oplus p'')\oplus q),
\ee
as is natural from a physical point of view. In this case, the problems related to nonassociativity are more evident, since one may choose
for example $\cK^\uno=(p\oplus q)-(p'\oplus(p''\oplus q))$, but our choice seems to be the most natural, at least at the classical level.
However, for more complicated processes, this simple prescription may not be consistent with the principle of relativity (see next section),
and one may be forced to consider several different choices compatible with the nonassociative and noncommutative character of the composition
law. Again, one may interpret the alternative choices as different channels for the interaction.

As in the previous section, it can be checked that all the \eom are invariant under translations generated by
the choice (\ref{cl}) of the conservation laws.
In fact, the time derivative of the \coo are unchanged under translations, because the nontrivial terms in the \trans depend only
on the conserved momenta.
Moreover, substituting (\ref{bc}), (\ref{bc2}) into (\ref{trans}) and (\ref{trans2}), one can see that also the \bc maintain their form if
$\x_{\uno B}^\m-\x_{\uno A}^\m=b^\m$ and (\ref{cond}) hold.

It must be noted that both the introduction of $q$ in $\cK^\uno$ and the ordering of
the momenta are rather ad hoc assumptions and are justified by the fact that these
conservation laws give the correct result. A drawback is that they give rise to a sort of entanglement, since with these prescriptions
the interaction at a point depends on the past history of the particle (this recalls the spectator problem in DSR).

It appears therefore that in the Snyder case one can obtain a relativistic description of the interaction between particles using the same
prescriptions for the interaction terms as in the \kp case. It is likely that this recipe is valid for any relative locality model.

\section{Time delay and dual curvature lensing in Snyder spacetime with interactions}
\label{sec:timedelay}

We want to consider now, in our Snyder framework, a typical process
suitable to highlight the effects of symmetry deformations on the
time delay in the detection of ultra-high energy particles emitted
by astrophysical sources. These effects have been proved to be of
phenomenological relevance within the most studied relative-locality
scenario with $\kappa$-Poincaré momentum space~\cite{anatomy},
as for instance for the (in-vacuo) propagation of gamma ray bursts (GRB) or astrophysical
neutrinos (see~\cite{IceCubeNat} for an up-to-date analysis), providing
an example where the Planck-scale effects manifest themselves in a
linear dependence of the times of arrival on the energies of the detected
particles.

In the case of non-interacting particles, time delay effects have
been investigated for the Snyder model in ref.~\cite{MS}, where
it has been shown that they are absent, due to the preservation of
the linear action of the Lorentz transformations on phase space. However,
in the case of interacting particles, the composition law of momenta
is deformed \cite{BM,IMS}, and time-delay effects may arise.

Besides time delays, another kind of effect, involving the direction
of particle propagation, has been studied in the relative locality
literature~\cite{FreSmolLens,spinning,trans,rhoMink},
where it has been named ``dual gravity'' or, more appropriately\footnote{Since the effect is due only to the curvature of momentum space, and
does not depend on the dynamics of geometry, the latter choice appears more appropriate \cite{rhoMink}.}, ``dual curvature'' lensing.
The effect is similar in its manifestation to the more standard gravitational lensing: in both cases the angle of observation of signals produced by some astrophysical sources is deflected so that their apparent position is different from their true position.
But, while in the case of gravitational lensing this is due to the bending of light by massive objects between the source and the observer, dual curvature lensing is caused only by the properties of curvature of momentum space characterizing particle kinematics, and thus occurs also in in-vacuo propagation.
As for in-vacuo dispersion, frameworks that provide a dual curvature lensing depending linearly on the propagating particle energies can be tested with presently available experimental observations~\cite{IceCubeLens}.
We will set the framework of our analysis
so to be able to take into account also this kind of effects.

\begin{figure}[h!]
\centering \includegraphics[scale=0.7]{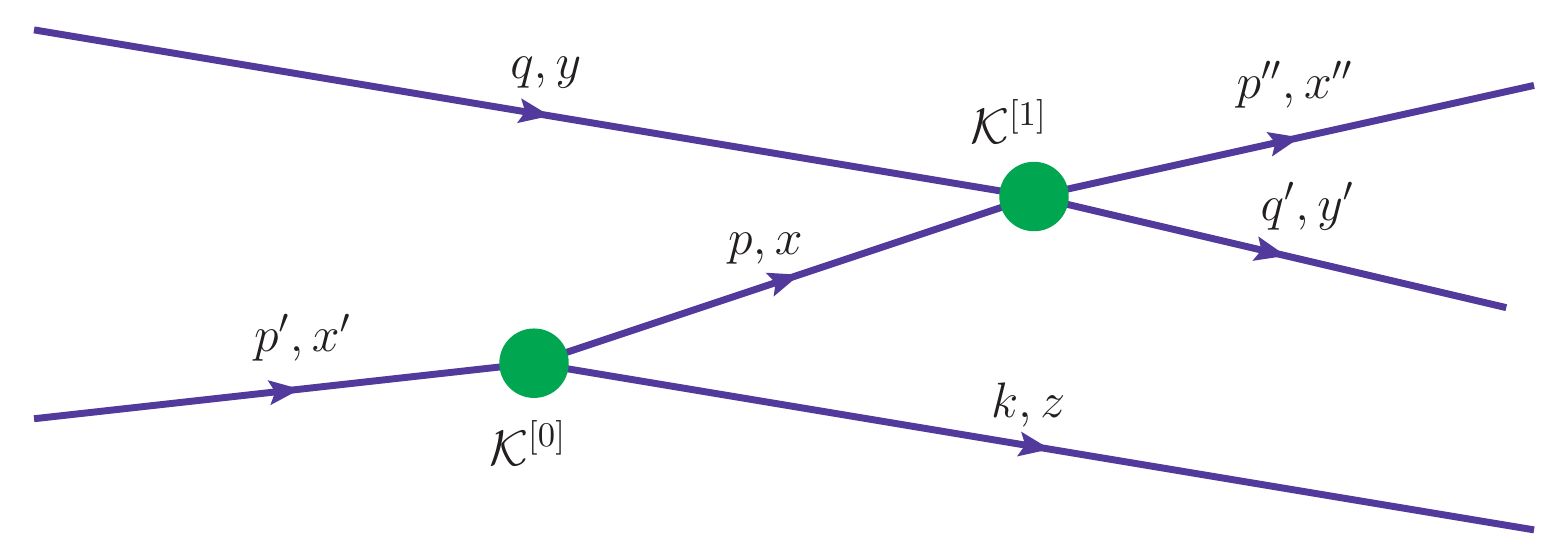}\caption{\label{TimeDelayFig1}\small
The chain of processes here schematically depicted can be interpreted as a high energy particle $(p',x')$ (e.g.~a highly boosted pion) decaying at $s_0$ (the source) into a particle $(k,z)$ propagating freely, and a second particle $(p,x)$ (a GRB-photon or a GRB-neutrino) interacting at $s_1$ (the detector) with a particle $(q,y)$ and finally producing $(p'',x'')$ and $(q',y')$.
}
\end{figure}

We consider the process depicted in Fig.~\ref{TimeDelayFig1}. This
process can be used to describe schematically both the GRB-photon
and GRB-neutrino scenarios. In the first case the particle $\left(p',x'\right)$
can be interpreted as a highly boosted neutral pion, which decays
at the source (at $s_{0}$) into a ``hard'' (high energy) GRB photon
$(p,x)$ and a second photon $(k,z)$. The GRB photon $(p,x)$ propagates
freely and is detected through its interaction at  $s_{1}$
with a particle $(q,y)$ (for instance in the excitation and de-excitation
of an atom of the detector) with which it exchanges a small amount
of its momentum. In the second case the particle
$\left(p',x'\right)$ can be interpreted as a highly boosted charged
pion, that decays (at $s_{0}$) into a muon $(k,z)$ and a muon-neutrino
$(p,x)$. The latter propagates freely until it is detected through its
interaction with a particle $(q,y)$ (for instance a deep inelastic
scattering with a nucleon of an atom at the detector) at the detector
at $s_{1}$.

The process of Fig.~\ref{TimeDelayFig1} can be described by an action
of the form of (\ref{actionCausal}), where the interactions are
encoded in the boundary terms
\begin{equation}
\begin{gathered}{\cal K}^{[0]}=\left(q\oplus p'\right)-\left(q\oplus p\right)\oplus k,\\
{\cal K}^{[1]}=\left(q\oplus p\right)\oplus k-\left(p''\oplus q'\right)\oplus k.
\end{gathered}
\label{boundary}
\end{equation}
Notice that in this case it is not possible to respect the naturalness
condition stated in the previous section. In fact, the request or
relativity, eq.~(\ref{rel}), forces one to choose the same ordering
for the double sum $q\oplus p\oplus k$ in the two interaction terms,
preventing the definition of any intuitive criterion for this choice.
Due to the nonassociativity, different orderings are not equivalent,
and then one can define alternative interaction terms, as for example
\begin{equation}
\begin{gathered}{\cal K}^{[0]}=\left(q\oplus p'\right)-q\oplus\left(p\oplus k\right),\\
{\cal K}^{[1]}=q\oplus\left(p\oplus k\right)-\left(p''\oplus q'\right)\oplus k,
\end{gathered}
\label{alter}
\end{equation}
or even further ones obtained by permutations of the momenta. These
alternative choices would give rise to results analogous to those
following from (\ref{boundary}), but with different numerical coefficients.
Again, these could be considered as different channels of the interaction.

For simplicity, we shall restrict our analysis to (2+1) dimensions, which turns out to
be sufficient for our discussion, and consider only terms up to first
order in $\beta\sim1/M_{pl}^{2}$.

We want to compare the arrival time and direction of the hard GRB-photon
(or the GRB-neutrino) $(p,x)$ at the detector with the arrival time
(and direction) of a second ``soft'' photon $(p_{t},x_{t})$, of
much lower energy, emitted at the source simultaneously to the hard
$(p,x)$ photon, but in an uncorrelated process. We consider a first
observer $A$ local to the emission event $s_{0}$ and a second observer
$B$, at rest with respect to $A$, local to the detector. Assuming
the distance\footnote{The notion of distance here is subtle, since it needs to be
operatively defined, for instance by considering the actual exchange
of signals between the two observers at the emission and at the detector,
so that its definition may be affected by the non-standard spacetime
framework. Here we refer to the notion of distance one would have
in ordinary special relativity. This assumption will be justified
in the course of the analysis.} between the emission event and the detector to be $T$, we define
$B$'s coordinates to be obtained from $A$'s coordinates by a pure
translation with parameters
\begin{equation}
b_{\mu}\equiv\left(b_{0},b_{1},b_{2}\right)=\left(T,T,0\right).\label{TransParam}
\end{equation}
This amounts to define the origin of $B$'s frame at what would be
the point of the detector reached by a standard special relativistic
photon emitted at $A$'s in the direction of its $x_{1}^{A}$ axis.

\begin{figure}[h!]
\centering \includegraphics[scale=0.7]{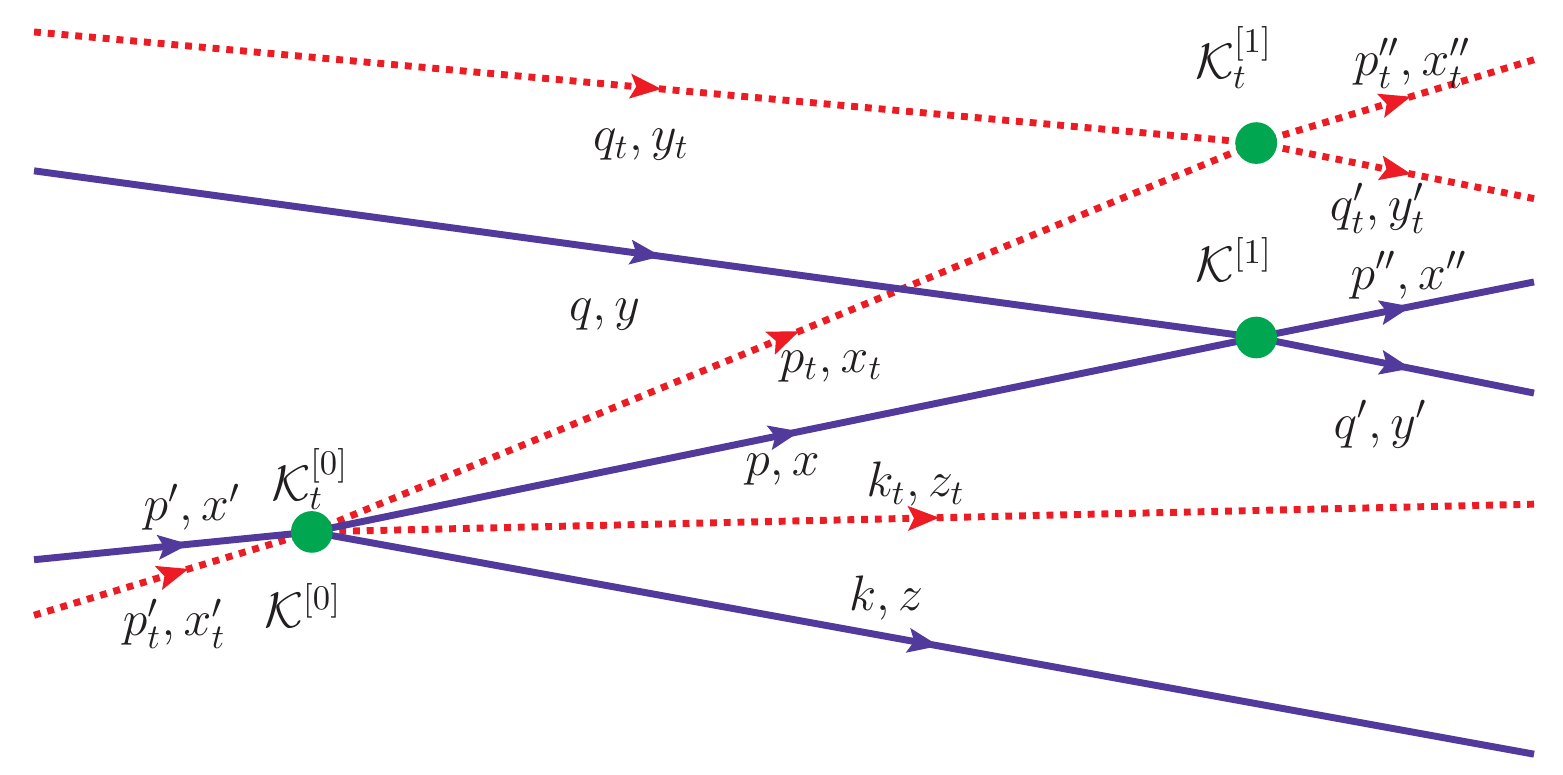}\caption{\small In this figure we schematically represent two uncorrelated processes of the same kind of Fig.~\ref{TimeDelayFig1}, one describing a lower energy process (red, dashed) characterizing the detection of a trigger (GRB) photon with respect to which the time and angle of detection of the higher energy particle (GRB-photon/GRB-neutrino) $(p,x)$, described by the second (blue, solid) higher energy process, is predicted.
\label{TimeDelayFig2}}
\end{figure}

We assume now that the ``trigger'' photon $(p_{t},x_{t})$ is emitted
at the source in a process similar to the one of Fig.~\ref{TimeDelayFig1},
but not correlated to the former. We can picture schematically the
two processes as in Fig.~\ref{TimeDelayFig2}. We will first analyze
the process of Fig.~\ref{TimeDelayFig1} generically and then impose
the energy conditions on the particles specifying the high (hard GRB
photon/GRB-neutrino) and low energy (trigger) processes.

Integrating with respect to $s$ Eqs. (\ref{eomp}) (or (\ref{eom})), we see that
$A$ and $B$ describe a generic particle $(p,x)$ to travel along
the worldlines
\begin{equation}
\begin{split}x_{A,B}^{j}\left(x_{A,B}^{0}\right) & =\bar{x}_{A,B}^{j}+\frac{\dot{x}_{A,B}^{j}}{\dot{x}_{A,B}^{0}}\Bigg|_{p_{0}=p_{0}\left(p_{1}\right)}\left(x_{A,B}^{0}-\bar{x}_{A,B}^{0}\right)\\
 & =\bar{x}_{A,B}^{j}+\frac{p^{j}}{p^{0}\left(\vec{p}\right)}\left(x_{A,B}^{0}-\bar{x}_{A,B}^{0}\right),
\end{split}
\label{EqMotionAB}
\end{equation}
where $\bar{x}_{A,B}^{\mu}\equiv(\bar{x}_{A,B}^{0},\bar{x}_{A,B}^{1},\bar{x}_{A,B}^{2})$
are constants of motion specifying the worldline initial conditions,
$p^{0}\left(\vec{p}\right)=\sqrt{\vec{p}^{2}+m^{2}}$ is the on-shell
relation ($\vec{p}\equiv(p_{1},p_{2})$), and the momentum of the
particle is identical for observers connected by a pure translation.
With the definition given in the previous sections, translations act
rigidly on the worldline coordinates so that each point of the worldline
changes by the the same amount. If the relation between the coordinates
of $A$ and $B$ after a translation is given by
\begin{equation}
x_{B}^{\mu}=x_{A}^{\mu}+\delta x^{\mu},\label{translation}
\end{equation}
using the boundary terms (\ref{boundary}) the shifts $\delta x^{\mu}$ are given by the Poisson brackets:
\begin{equation}
\delta x^{\mu}=b^{\nu}\left\{ \left(\left(q\oplus p\right)\oplus k\right)_{\nu},x^{\mu}\right\} ,
\end{equation}
where (see (\ref{sny}))
\begin{equation}
\left\{ p_{\mu},x^{\nu}\right\} =-\delta_{\mu}^{\nu}+\beta p_{\mu}p^{\nu},\qquad\left\{ q_{\mu},x^{\nu}\right\} =\left\{ k_{\mu},x^{\nu}\right\} =0.
\end{equation}

From (\ref{Oplus}) we get, at first order in $\beta$,
\begin{equation}
\left(q\oplus p\right)_{\mu}\simeq q_{\mu}+p_{\mu}-\frac{\beta}{2}\left(q\bdot p\,q_{\mu}+q\bdot q\,p_{\mu}+2q\bdot p\,p_{\mu}\right),
\end{equation}
and
\begin{equation}
\begin{split}\left(\left(q\oplus p\right)\oplus k\right)_{\mu}\simeq & \left(q\oplus p\right)_{\mu}+k_{\mu}-\frac{\beta}{2}\left(\left(q\oplus p\right)\bdot k\left(q\oplus p\right)_{\mu}+\left(q\oplus p\right)\bdot
\left(q\oplus p\right) k_{\mu}+2\left(q\oplus p\right)\bdot k\,k_{\mu}\right)\\
\simeq & \ q_{\mu}+p_{\mu}+k_{\mu}-\frac{\beta}{2}\left(q\bdot p+q\bdot k+p\bdot k\right)q_{\mu}\\
 & -\frac{\beta}{2}\left(q\bdot q+2q\bdot p+q\bdot k+p\bdot k\right)p_{\mu}-\frac{\beta}{2}\left(q\bdot q+p\bdot p+2q\bdot p+2q\bdot k+2p\bdot k\right)k_{\mu},
\end{split}
\end{equation}
so that we find
\begin{equation}
\begin{split}\delta x^{\mu}\simeq & -b^{\mu}+\frac{\beta}{2}\left(q\bdot q+2q\bdot p+q\bdot k+p\bdot k\right)b^{\mu}+\beta\left(b\bdot p+b\bdot k\right)p^{\mu}\\
 & +\frac{\beta}{2}\left(b\bdot q+2b\bdot p+2b\bdot k\right)q^{\mu}+\frac{\beta}{2}\left(b\bdot q+b\bdot p+2b\bdot k\right)k^{\mu} .
\end{split}
\label{xTranslation}
\end{equation}

We now specify the energy conditions for both processes. Consider first
the process for the trigger photon. This is described by the above
equations by setting $p,q,k\rightarrow p_{t},q_{t},k_{t}$. In order
to take into account possible dual curvature effects~\cite{spinning},
we allow the photon to be emitted at $A$ with a (small) angle $\alpha_{t}$
between the $x_{A}^{1}$ and $x_{A}^{2}$ axis, to be determined by
the condition that the photon is detected at $B'$s spatial origin
$\vec{\bar{x}}_{B}=(\bar{x}_{B}^{1},\bar{x}_{B}^{2})=\left(0,0\right)$.
The presence of this angle can be implemented by defining the photon
momentum as\footnote{Beware that $p_t^2,k_t^2,q_{t}^{2}$ represent
the second component of the momentum vectors, and not the squares
of $p_t,k_t,q_{t}$, to which we reserve the notation $p_t\cdot p_t, q_t\cdot q_t, k_t\cdot k_t$.}
\begin{equation}
p_{t}^{\mu}\equiv\left(p_{t}^{0},p_{t}^{1},p_{t}^{2}\right)=\left(\left|\vec{p}_{t}\right|,\left|\vec{p}_{t}\right|\cos\alpha_{t},\left|\vec{p}_{t}\right|\sin\alpha_{t}\right).\label{momentumTrig}
\end{equation}

In $A$'s frame the equations of motion for the photon become
\begin{equation}
\left(x_{t}^{1}\right)_{A}=\left(x_{t}^{0}\right)_{A}\cos\alpha_{t},\qquad\left(x_{t}^{2}\right)_{A}=\left(x_{t}^{0}\right)_{A}\sin\alpha_{t},\label{worldlineAt}
\end{equation}
where we have imposed the initial conditions $\bar{x}_{A}^{\mu}\equiv(0,0,0)$
enforcing the photon to be emitted at $A$'s spacetime origin. Imposing
$\vec{\bar{x}}_{B}=\left(0,0\right)$ in (\ref{EqMotionAB}) we find
the equations of motion in $B$'s frame
\begin{equation}\label{worldlineB}
\left(x_{t}^{1}\right)_{B}=\left[\left(x_{t}^{0}\right)_{B}-\left(\bar{x}_{t}^{0}\right)_{B}\right]\cos\alpha_{t},\qquad
\left(x_{t}^{2}\right)_{B}=\left[\left(x_{t}^{0}\right)_{B}-\left(\bar{x}_{t}^{0}\right)_{B}\right]\sin\alpha_{t}.
\end{equation}

Substituting~(\ref{translation}) in~(\ref{worldlineB}), and using (\ref{worldlineAt}), we get
\begin{equation}
\tan\alpha_{t}=\frac{\delta x_{t}^{2}}{\delta x_{t}^{1}},\qquad\left(\bar{x}_{t}^{0}\right)_{B}=\delta x_{t}^{0}-\delta x_{t}^{1}\sqrt{1+\tan^{2}\alpha_{t}}.\label{angleDelayt1}
\end{equation}
It follows from (\ref{xTranslation}) and (\ref{TransParam}) that
\begin{equation}
\begin{split}\alpha_{t}\simeq & \tan\alpha_{t}\simeq-\beta\left[\left(p^{0}-p^{1}\right)+\left(k^{0}-k^{1}\right)\right]p^{2}\\
 & -\frac{\beta}{2}\left[\left(q^{0}-q^{1}\right)+\left(p^{0}-p^{1}\right)+2\left(k^{0}-k^{1}\right)\right]k^{2}\\
 & -\frac{\beta}{2}\left[\left(q^{0}-q^{1}\right)+2\left(p^{0}-p^{1}\right)+2\left(k^{0}-k^{1}\right)\right]q^{2},
\end{split}
\label{alphat1}
\end{equation}
where  we are neglecting terms of order $\beta^{2}$.
Notice now that, since the angle $\alpha_{t}$ is of order $\beta$,
as can be seen from (\ref{momentumTrig}), the difference
between $p^{0}$ and $p^{1}$ can be neglected in (\ref{alphat1}).
Similarly, assuming\footnote{The pion $(p',x')$ decaying at the source is highly boosted, so that the product particles $\left(p,x\right)$ and ($k,z$) can be taken
to be collinear. We keep this assumption also for the trigger photon,
which we consider also to be a GRB photon (of lower energy).} the direction of the $\left(k_{t},z_{t}\right)$ worldline to be
collinear to the $\left(p_{t},x_{t}\right)$ one, so that $k_{t}^{\mu}\equiv(|\vec{k}_{t}|,|\vec{k}_{t}|\cos\alpha,|\vec{k}_{t}|\sin\alpha)$,
then $k_{t}^{0}-k_{t}^{1}\simeq O\left(\beta\right)$. Moreover $p_t^2 \simeq \alpha_t|\vec{p}_{t}| = O(\beta)$ as well as $k_t^2 \simeq \alpha_t|\vec{k}_{t}| = O(\beta)$ and, at linear order in $\beta$, we are left with
\begin{equation}
\alpha_{t}\simeq-\frac{\beta}{2}\left(q_{t}^{0}-q_{t}^{1}\right)q_{t}^{2}.\label{alphat}
\end{equation}

Before discussing the interpretation of this angle, let us calculate
the time of detection of the trigger photon, given by the second of
equations (\ref{angleDelayt1}). Using again (\ref{xTranslation})
and (\ref{TransParam}) we find
\begin{equation}
\begin{split}\left(\bar{x}_{t}^{0}\right)_{B}\simeq & \beta T\left[\left(p_{t}^{0}-p_{t}^{1}\right)+\left(k_{t}^{0}-k_{t}^{1}\right)\right]\left(p_{t}^{0}-p_{t}^{1}\right)\\
 & +\frac{\beta}{2}T\left[2\left(p_{t}^{0}-p_{t}^{1}\right)+\left(q_{t}^{0}-q_{t}^{1}\right)+2\left(k_{t}^{0}-k_{t}^{1}\right)\right]\left(q_{t}^{0}-q_{t}^{1}\right)\\
 & +\frac{\beta}{2}T\left[\left(p_{t}^{0}-p_{t}^{1}\right)+\left(q_{t}^{0}-q_{t}^{1}\right)+2\left(k_{t}^{0}-k_{t}^{1}\right)\right]\left(k_{t}^{0}-k_{t}^{1}\right).
\end{split}
\end{equation}
The same considerations that led us to (\ref{alphat}) reduce the
last equation to
\begin{equation}
\left(\bar{x}_{t}^{0}\right)_{B}\simeq\frac{\beta}{2}T\left(q_{t}^{0}-q_{t}^{1}\right)^{2}.\label{Delayt}
\end{equation}

It emerges from Eqs. (\ref{alphat}) and (\ref{Delayt}) that
the angle and time with which the trigger photon is detected at $B$
depend on the details of the interaction at the detector. In particular
we can set
\begin{equation}
q_{t}^{\mu}\equiv\left(E_{q_{t}},\left|\vec{q}_{t}\right|\cos\theta_{t},\left|\vec{q}_{t}\right|\sin\theta_{t}\right)
\end{equation}
and obtain
\begin{equation}
\alpha_{t}\simeq-\frac{\beta}{2}\left(E_{q_{t}}-\left|\vec{q}_{t}\right|\cos\theta_{t}\right)\left|\vec{q}_{t}\right|\sin\theta_{t},\qquad\left(\bar{x}_{t}^{0}\right)_{B}\simeq\frac{\beta}{2}T\left(E_{q_{t}}-\left|\vec{q}_{t}\right|\cos\theta_{t}\right)^{2}.\label{angleDelayt}
\end{equation}

We can repeat exactly the same procedure that has led us to (\ref{angleDelayt})
to derive the angle $\alpha$ and the arrival time for the hard GRB-photon
(or the GRB-neutrino) ($p,x$), and find
\begin{equation}
\alpha\simeq-\frac{\beta}{2}\left(E_{q}-\left|\vec{q}\right|\cos\theta\right)\left|\vec{q}\right|\sin\theta,\qquad\bar{x}_{B}^{0}\simeq\frac{\beta}{2}T\left(E_{q}-\left|\vec{q}\right|\cos\theta\right)^{2}.\label{angleDelayh}
\end{equation}
We thus finally find that the hard particle (the hard GRB-photon or
the GRB-neutrino) is detected, with respect to the trigger photon, at an
angle and time delay
\begin{equation}
\begin{gathered}\Delta\alpha\simeq-\frac{\beta}{2}\left(\left(E_{q}-\left|\vec{q}\right|\cos\theta\right)\left|\vec{q}\right|\sin\theta-\left(E_{q_{t}}-\left|\vec{q}_{t}\right|\cos\theta_{t}\right)\left|\vec{q}_{t}\right|\sin\theta_{t}\right),\\
\Delta t\simeq\frac{\beta}{2}T\left(\left(E_{q}-\left|\vec{q}\right|\cos\theta\right)^{2}-\left(E_{q_{t}}-\left|\vec{q}_{t}\right|\cos\theta_{t}\right)^{2}\right).
\end{gathered}
\label{angleDelay}
\end{equation}
If the particles $q_{t}$ and $q$ at the detector are non relativistic and particularly such that their momentum is much smaller than their mass $(\left|\vec{q}\right|/m\ll1)$,
the last expressions reduce to
\begin{equation}
\begin{gathered}\Delta\alpha\simeq-\frac{\beta}{2}\left(m_{q}\left|\vec{q}\right|\sin\theta-m_{q_{t}}\left|\vec{q}_{t}\right|\sin\theta_{t}\right),\\
\Delta t\simeq\frac{\beta}{2}T\left(m_{q}^{2}-m_{q_{t}}^{2}\right).
\end{gathered}
\end{equation}

\begin{figure}[h!]
\begin{center}
\includegraphics[scale=0.7]{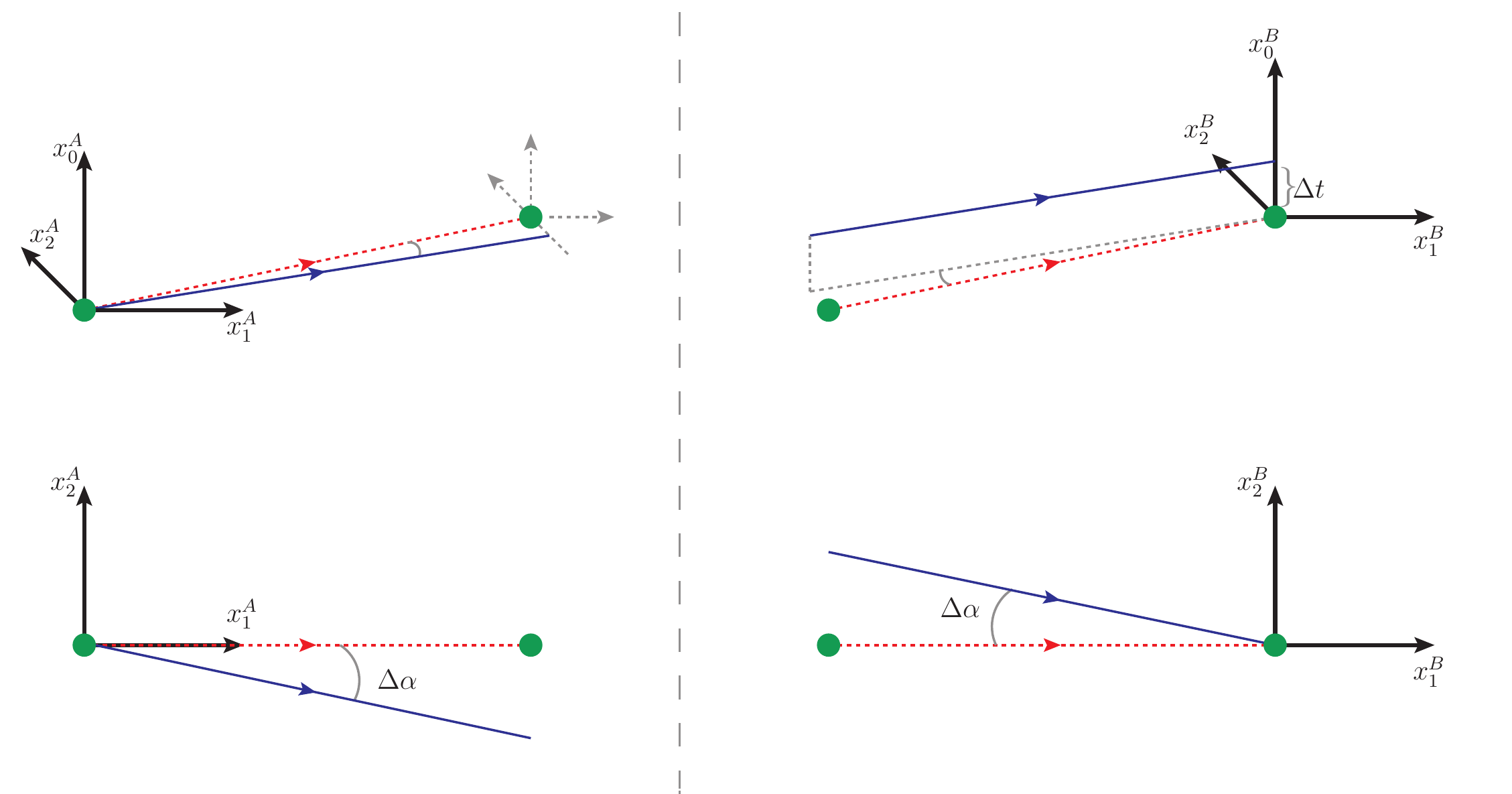}
\caption{
\label{FigDelayLensing}
\small We show here the joint effects of time delay
and dual curvature lensing on the hard GRB photon (or neutrino)
(blue, solid line) $\left(p,x\right)$ and the trigger photon (red, dashed line)
$(p_{t},x_{t})$. The left panel represents $A$'s frame, the right
panel $B$'s frame. We assume for simplicity that $m_{q}\gg m_{q_{t}}$,
so that the trigger photon is detected approximatively in $B$'s spacetime
origin and with $\alpha_{t}\sim0$.}
\end{center}
\end{figure}

The results are pictured in Fig. \ref{FigDelayLensing}.

Equation (\ref{angleDelay}) shows that both time-delay and dual-curvature
lensing effects are present, depending on the details of
the interaction between the ($p,x$) (or ($p_{t},x_{t}$)) particle
and the detector. The time-delay effect is extremely tiny, since it is proportional
to the square of the ratio between the particle masses at the detector,
like for instance atoms or nucleons composing the detector, and the
Planck mass ($\beta\propto1/M_{pl}^{2}$), if the Snyder deformation
has to be understood as generated by some quantum gravity effect.
The only amplifying factor for the time delay is the distance $T$
traveled by the photon from the source to the detector. Notice however
that the effect does not depend on the photon (or neutrino) energies.
This means that in principle the time delay induced by the Snyder deformation
could be investigated considering also low-energy particles, the drawback
being of course that the particle energy does not act as an amplifier
for the Planckian effect. Thus the effect is far beyond the reach
of present astrophysical experiments, but our analysis shows that,
at least within the relative-locality scenario, one could in principle
devise a detector capable of testing a deformation of spacetime symmetries
of Snyder type. Similar considerations hold for the lensing effect, but
this is much fainter, and has only a theoretical relevance.

An important remark is that the effects we have found mostly depend
on the properties of the detector, rather than of the incoming particles.
This is a distinctive feature of the Snyder phenomenology.

Finally, it is interesting to see if the same effects are obtained also for a
different choice of the interaction term, like (\ref{alter}), instead
of (\ref{boundary}). In this case, one has
\begin{equation}
\begin{split}\left(q\oplus\left(p\oplus k\right)\right)_{\mu}\simeq & \ q_{\mu}+p_{\mu}+k_{\mu}-\frac{\beta}{2}\left(q\bdot p+q\bdot k\right)q_{\mu}\\
 & -\frac{\beta}{2}\left(q\bdot q+2q\bdot p+2q\bdot k+p\bdot k\right)p_{\mu}-\frac{\beta}{2}\left(q\bdot q+p\bdot p+2q\bdot p+2q\bdot k+2p\bdot k\right)k_{\mu},
\end{split}
\end{equation}
and (\ref{xTranslation}) becomes
\begin{equation}
\begin{split}\delta x^{\mu}\simeq & -b^{\mu}+\frac{\beta}{2}\left(q\bdot q+2q\bdot p+2q\bdot k+p\bdot k\right)b^{\mu}+\beta\left(b\bdot p+b\bdot k\right)p^{\mu}\\
 & +\frac{\beta}{2}\left(b\bdot q+2b\bdot p+2b\bdot k\right)q^{\mu}+\frac{\beta}{2}\left(b\bdot p+2b\bdot k\right)k^{\mu}.
\end{split}
\end{equation}
Repeating the same steps as before, the calculation of the time delay
$\D t$ and of the dual curvature lensing $\D\alpha$ again reproduces
the result (\ref{angleDelay}). At this order of approximation, the nonassociativity
is therefore not relevant for the experimental predictions.

Due to the noncommutativity, one may however also modify the ordering
of the momenta in (\ref{boundary}) or (\ref{alter}). In this case,
a factor of 2 may appear in (\ref{angleDelay}) for some permutations,
but the qualitative results are not modified. It is reasonable to
assume that an actual measurement would average among all the possible
outcomes predicted by modifying the interaction term.

\section{Discussion}
In this paper we have extended the investigation of the dynamics of relative locality to
the case of the Snyder model. Formally, this generalization does not introduce new
features in comparison with previously studied models, except for the nonassociativity of
the momenta addition law, that entails some ambiguities in the definition of the interaction
terms. However, phenomenological prediction can be rather different.

More specifically, we have found that in the relative locality framework Snyder momentum
space predicts, under certain conditions, a non-null time delay in
the arrival of photons emitted simultaneously from a distant source.
It is worth comparing the leading term characterizing the time delay
effect for the Snyder case with the one for the $\kappa$-Poincaré
case obtained in~\cite{anatomy}:
\begin{equation}
\text{Snyder}:\Delta t_{S}\simeq T_{\gamma}\frac{\Delta m_{\text{det}}^{2}}{E_{p}^{2}},\qquad\kappa\text{-Poincaré}:\Delta t_{\kappa}\simeq T_{\gamma}\frac{\Delta E_{\gamma}}{E_{p}}.\label{comparison}
\end{equation}
In both instances the effect depends on the distance (in time) $T_{\gamma}$
traveled by the photons from the source to the detector. However,
for the Snyder case it does not depend on the photon energies $E_{\gamma}$,
as expected, since in Snyder momentum space the on-shell relation
for a massless particle is undeformed (and thus there is no in-vacuo
dispersion for photons) so that, besides being of second order in
$1/E_{p}$, it lacks one of the two sources of amplification that
balance the smallness of the Planckian effect. As a matter of fact,
the Snyder effect is of a different nature with respect to the $\kappa$-Poincaré
one, since it depends on the mass difference $\Delta m_{\text{det}}$
of the particles at the detector with which two photons emitted simultaneously
at the source respectively interact, rather than on the energy difference
$\Delta E_{\gamma}$ between the traveling photons, as for the $\kappa$-Poincaré
case. This means that the effect we have found for the Snyder model
affects the propagation of photons independently of their energies,
but only depending on the details of the processes by which they interact
at the detector, so that it could be in principle investigated using
astrophysical events emitting low-energy photons. Obviously the effect
is too tiny to be realistically taken into consideration experimentally.

We notice that an effect similar to the one we found
for Snyder, with the time delay driven by the ratio $\Delta m_{\text{det}}/E_{p}$,
is present also for the $\kappa$-Poincaré case, even if it is subleading
with respect to the term in~(\ref{comparison}). Indeed, if one repeats
the analysis of time delay in the framework of~\cite{anatomy} (we
skip the details of the derivation and leave it to the reader), and
sets the energy conditions we used in Sec.~\ref{sec:timedelay},
one obtains the time delay
\begin{equation}
\Delta t_{\kappa}\simeq T_{\gamma}\frac{\Delta E_{\gamma}}{E_{p}}+T_{\gamma}\frac{\Delta m_{\text{det}}}{E_{p}}.
\end{equation}
We understand this effect as a feature of the formalism of relative
locality, where the non-trivial summation law of the momenta of the
particles entering the interaction manifests itself in a dependence
of the detection times on the kinematical details of the process.

We have also found that in Snyder framework an effect of dual curvature
lensing~\cite{FreSmolLens,spinning,trans,rhoMink} is present. The effect is similar
to the one discussed in discussed in~\cite{spinning} for a
different framework\footnote{In the case studied in~\cite{spinning} only the dual curvature
effect is present, while time delay is absent.}, where the magnitude of the corrections depends only on the details of the interaction,
and not on the energy of the propagating particles. Again the effect is extremely
tiny, since not even the propagation distance $T$ can provide a source
of amplification, and has therefore only a theoretical relevance.

To conclude, we recall that the Snyder model differs from \kp models in several respects:
it preserves the linear action of the Lorentz group and by consequence the leading correction to special relativity are of order $1/E_p^2$ \cite{IMS},
and moreover the law of addition of momenta is nonassociative. The former properties imply that relative locality effects can only be detected
in many-particle interactions and are highly suppressed with respect to the corresponding \kp corrections.
The latter one gives rise to much more involved problems in the definition of the interactions, but does not affect phenomenology significantly.


\newpage

\begin{thebibliography}{0}

\bibitem{GACphen}
  G.~Amelino-Camelia,
  Living Rev.\ Rel.\  {\bf 16} (2013) 5
  [arXiv:0806.0339 [gr-qc]].

\bibitem{MattinglyRev}
  D.~Mattingly,
  Living Rev.\ Rel.\  {\bf 8} (2005) 5
  [gr-qc/0502097].

\bibitem{GACGRB}
  G.~Amelino-Camelia, J.~R.~Ellis, N.~E.~Mavromatos, D.~V.~Nanopoulos and S.~Sarkar,
  Nature {\bf 393} (1998) 763
  [astro-ph/9712103].

\bibitem{AlfaroNeutrinos}
  J.~Alfaro, H.~A.~Morales-Tecotl and L.~F.~Urrutia,
  Phys.\ Rev.\ Lett.\  {\bf 84} (2000) 2318
  [gr-qc/9909079].

\bibitem{JacobPiran}
  U.~Jacob and T.~Piran,
  Nature Phys.\  {\bf 3} (2007) 87
  [hep-ph/0607145].

\bibitem{IceCubeNat}
  G.~Amelino-Camelia, G.~D'Amico, G.~Rosati and N.~Loret,
  Nat.\ Astron.\  {\bf 1} (2017) 0139
  [arXiv:1612.02765 [astro-ph.HE]].

\bibitem{GACJrFioreGRB}
  G.~Amelino-Camelia, G.~D'Amico, F.~Fiore, S.~Puccetti and M.~Ronco,
  arXiv:1707.02413 [astro-ph.HE].

\bibitem{GACDSR}
  G.~Amelino-Camelia,
  Int.\ J.\ Mod.\ Phys.\ D {\bf 11} (2002) 35
  [gr-qc/0012051];
  Phys.\ Lett.\ B {\bf 510} (2001) 255
  [hep-th/0012238].

\bibitem{JurekDSR}
  J.~Kowalski-Glikman,
  Phys.\ Lett.\ A {\bf 286} (2001) 391
 [hep-th/0102098].

\bibitem{SmolMagDSR}
  J.~Magueijo and L.~Smolin,
  Phys.\ Rev.\ D {\bf 67} (2003) 044017
  [gr-qc/0207085].

\bibitem{DSRFRW}
  G.~Rosati, G.~Amelino-Camelia, A.~Marciano and M.~Matassa,
  Phys.\ Rev.\ D {\bf 92} (2015) 124042
  [arXiv:1507.02056 [hep-th]].

\bibitem{DualLensing}
  G.~Amelino-Camelia, L.~Barcaroli, S.~Bianco and L.~Pensato,
  Adv.\ High Energy Phys.\  {\bf 2017} (2017) 6075920
  [arXiv:1708.02429 [gr-qc]].

\bibitem{RelLocPRL}
  G.~Amelino-Camelia, M.~Matassa, F.~Mercati and G.~Rosati,
  Phys.\ Rev.\ Lett.\  {\bf 106} (2011) 071301
  [arXiv:1006.2126 [gr-qc]].

\bibitem{kBob}
  G.~Amelino-Camelia, N.~Loret and G.~Rosati,
  Phys.\ Lett.\ B {\bf 700} (2011) 150
  [arXiv:1102.4637 [hep-th]].

\bibitem{RelLocPrinciple}
  G.~Amelino-Camelia, L.~Freidel, J.~Kowalski-Glikman and L.~Smolin,
  Phys.\ Rev.\ D {\bf 84} (2011) 084010
  [arXiv:1101.0931 [hep-th]];
  Gen.\ Rel.\ Grav.\  {\bf 43} (2011) 2547
  [arXiv:1106.0313 [hep-th]].

\bibitem{CarmonaRelLoc}
  J.~M.~Carmona, J.~L.~Cortes, D.~Mazon and F.~Mercati,
  Phys.\ Rev.\ D {\bf 84} (2011) 085010
  [arXiv:1107.0939 [hep-th]].

\bibitem{JurekReview}
  J.~Kowalski-Glikman,
  Int.\ J.\ Mod.\ Phys.\ A {\bf 28} (2013) 1330014
  [arXiv:1303.0195 [hep-th]].

\bibitem{Palmisano}
  G.~Amelino-Camelia, G.~Gubitosi and G.~Palmisano,
  Int.\ J.\ Mod.\ Phys.\ D {\bf 25} (2016)  1650027
  [arXiv:1307.7988 [gr-qc]].

\bibitem{anatomy}
  G.~Amelino-Camelia, M.~Arzano, J.~Kowalski-Glikman, G.~Rosati and G.~Trevisan,
  Class.\ Quant.\ Grav.\  {\bf 29} (2012) 075007
  [arXiv:1107.1724 [hep-th]].

\bibitem{FlaGiukRelLoc}
  G.~Gubitosi and F.~Mercati,
  Class.\ Quant.\ Grav.\  {\bf 30} (2013) 145002
  [arXiv:1106.5710 [gr-qc]].

\bibitem{causality}
  G.~Amelino-Camelia, S.~Bianco, F.~Brighenti and R.~J.~Buonocore,
  Phys.\ Rev.\ D {\bf 91} (2015) 084045
  [arXiv:1401.7160 [gr-qc]].

\bibitem{multipart}
  J.~Kowalski-Glikman and G.~Rosati,
  Phys.\ Rev.\ D {\bf 91} (2015) 084061
  [arXiv:1412.0493 [hep-th]].


\bibitem{spinning}
  G.~Amelino-Camelia, M.~Arzano, S.~Bianco and R.~J.~Buonocore,
  Class.\ Quant.\ Grav.\  {\bf 30} (2013) 065012
  [arXiv:1210.7834 [hep-th]].

\bibitem{Snyder}
  H.~S.~Snyder,
  Phys.\ Rev.\  {\bf 71} (1947) 38.

\bibitem{Sn}
Y.~Gol'fand, JETP{\bf 10} (1960) 356; V.~Kadyshevskii, JETP{\bf 14} (1963) 1340;
G.~Jaroszkiewicz, \JoP{A28} (1995) L343;
J.~M.~Romero and A.~Zamora, \PR{D70} (2004) 105006 [hep-th/0408193];
R.~Banerjee, S.~Kulkarni and S.~Samanta, \JHEP{05} (2006) 077 [hep-th/0602151];
S.~Mignemi \IJMP{D24} (2015) 1550043  [arXiv:1308.0673 [hep-th]].

\bibitem{BM} M.~V.~Battisti and S.~Meljanac, \PR{D79} (2009) 067505 [arXiv:0812.3755 [hep-th]];
  \PR{D82} (2010) 024028 [arXiv:1003.2108 [hep-th]];
  F.~Girelli and E.~R.~Livine,
  JHEP {\bf 1103} (2011) 132
  [arXiv:1004.0621 [hep-th]].

\bibitem{MigSamSnyderRL}
  S.~Mignemi and A.~Samsarov,
  Phys.\ Lett.\ A {\bf 381} (2017) 1655
  [arXiv:1610.09692 [hep-th]].

\bibitem{AstutiSnyder}
  G.~Amelino-Camelia and V.~Astuti,
  Int.\ J.\ Mod.\ Phys.\ D {\bf 24} (2015) 1550073
  [arXiv:1404.4773 [hep-th]].

\bibitem{IMS}
  B.~Iveti\'c, S.~Mignemi and A.~Samsarov, \PR{D94} (2016) 064064
  [arXiv:1606.04692 1404.4773 [hep-th]].

\bibitem{IM}   B. Iveti\'c and S. Mignemi, arXiv:1711.07438 [hep-th].

\bibitem{MS}  S.~Mignemi and R.~\v Strajn, \PL{A380} (2016) 1714
 [arXiv:1509.05311 [hep-th]].

\bibitem{FreSmolLens}
  L.~Freidel and L.~Smolin,
  arXiv:1103.5626 [hep-th].

\bibitem{trans}
  G.~Amelino-Camelia, L.~Barcaroli and N.~Loret,
  Int.\ J.\ Theor.\ Phys.\  {\bf 51} (2012) 3359
  [arXiv:1107.3334 [hep-th]].

\bibitem{rhoMink}
  G.~Amelino-Camelia, L.~Barcaroli, S.~Bianco and L.~Pensato,
  Adv.\ High Energy Phys.\  {\bf 2017} (2017) 6075920
  [arXiv:1708.02429 [gr-qc]].

\bibitem{IceCubeLens}
  G.~Amelino-Camelia, L.~Barcaroli, G.~D'Amico, N.~Loret and G.~Rosati,
  Int.\ J.\ Mod.\ Phys.\ D {\bf 26} (2017) no.08,  1750076
  [arXiv:1609.03982 [gr-qc]].

\end{thebibliography}
\end{document}